\documentclass[draft]{agujournal2019}
\usepackage{apacite}
\usepackage{url} 
\usepackage{lineno}

\usepackage{graphicx}
\usepackage{array}
\usepackage{SIunits}
\usepackage{multirow}

\usepackage{xcolor}

\usepackage{ccaption}

\textbf{}
\draftfalse

\usepackage{booktabs}
\usepackage{tabularx}
\usepackage{tabulary}
\usepackage{numprint} 
\newcommand{\colortable}[1]{\begingroup\setlength{\fboxsep}{0pt}
\colorbox{lightgray}{#1}\endgroup}
\journalname{JGR: Space Physics}

\begin{document}

%
%



\title{Source of radio emissions induced by the Galilean moons Io, Europa and Ganymede: \textit{in situ} measurements by Juno}


%
%




\authors{C.~K.~Louis\affil{1,2,3}, P.~Louarn\affil{2},
B.~Collet\affil{4},
N.~Cl\'{e}ment\affil{2,5}, S.~Al~Saati\affil{2,6}, 
J.~R.~Szalay\affil{7}, V.~Hue\affil{8}, 
L.~Lamy\affil{3,4},
S.~Kotsiaros\affil{9}, W.~S.~Kurth\affil{10}, C.~M.~Jackman\affil{1},Y.~Wang\affil{2,11,12}, M.~Blanc\affil{2,5}, F.~Allegrini\affil{13,14}, J.~E.~P.~Connerney\affil{15}, D.~Gershman\affil{16}}


\affiliation{1}{School of Cosmic Physics, DIAS Dunsink Observatory, Dublin Institute for Advanced Studies, Dublin 15, Ireland}
\affiliation{2}{IRAP, Universit\'e de Toulouse, CNRS, CNES, UPS, Toulouse, France}
\affiliation{3}{LESIA, Observatoire de Paris, Universit\'e PSL, CNRS, Sorbonne Universit\'e, Universit\'e de Paris, Meudon, France}
\affiliation{4}{Pyth\'eas, Aix Marseille Universit\'e, CNRS, CNES, Marseille, France}
\affiliation{5}{Laboratoire d'Astrophysique de Bordeaux, Univ. Bordeaux, CNRS, B18N, All\'ee Geoffroy Saint--Hilaire, 33615 Pessac, France}
\affiliation{6}{CPHT, CNRS, Institut Polytechnique de Paris, Route de Saclay, 91128 Palaiseau, France}
\affiliation{7}{Department of Astrophysical Sciences, Princeton University, Princeton, New Jersey, USA}
\affiliation{8}{Aix--Marseille Universit\'{e}, CNRS, CNES, Institut Origines, LAM, Marseille, France}
\affiliation{9}{Technical University of Denmark: Kgs. Lyngby, Denmark}
\affiliation{10}{Department of Physics and Astronomy, University of Iowa, Iowa City, Iowa, USA}
\affiliation{11}{State Key Laboratory of Space Weather, National Space Science Center, Chinese Academy of Sciences, Beijing, China. }
\affiliation{12}{College of Earth and Planetary Sciences, University of Chinese Academy of Sciences, Beijing, China.}
\affiliation{13}{Southwest Research Institute, San Antonio, Texas, USA}
\affiliation{14}{Department of Physics and Astronomy, University of Texas at San Antonio, San Antonio, Texas, USA}
\affiliation{15}{Space Research Corporation, Annapolis, MD, USA 21403}
\affiliation{16}{NASA Goddard Space Flight Center, Greenbelt, MD, USA}





\correspondingauthor{C. K. Louis}{corentin.louis@dias.ie}




\begin{keypoints}
\item All Jupiter--moon radio emissions are shown to be similarly triggered by the CMI.
\item The crossed radio sources are colocated with either MAW, RAW or TEB footprints.
\item The crossed radio sources coincide with downward field--aligned currents and Alfv\'{e}n perturbations.
\end{keypoints}

%
%


\begin{abstract}
At Jupiter, part of the auroral radio emissions are induced by the Galilean moons Io, Europa and Ganymede. Until now, except for Ganymede, they have been only remotely detected, using ground--based radio--telescopes or electric antennas aboard spacecraft. The polar trajectory of the Juno orbiter allows the spacecraft to cross the range of magnetic flux tubes which sustain the various Jupiter--satellite interactions, and in turn to sample \textit{in situ} the associated radio emission regions.
In this study, we focus on the detection and the characterization of radio sources associated with Io, Europa and Ganymede. Using electric wave measurements or radio observations (Juno/Waves), \textit{in situ} electron measurements (Juno/JADE--E), and magnetic field measurements (Juno/MAG) we demonstrate that the Cyclotron Maser Instability (CMI) driven by a loss--cone electron distribution function is responsible for the encountered radio sources. We confirmed that radio emissions are associated with Main (MAW) or Reflected Alfv\'en Wing (RAW), but also show that for Europa and Ganymede, induced radio emissions are associated with Transhemispheric Electron Beam (TEB). For each traversed radio source, we determine the latitudinal extension, the CMI--resonant electron energy, and the bandwidth of the emission. We show that the presence of Alfv\'{e}n perturbations and downward field aligned currents are necessary for the radio emissions to be amplified.
\end{abstract}

%
%

%


%
%
%
%

\section{Introduction}
\label{sec:intro}

One of the main objectives of the Juno mission is to probe Jupiter's auroral regions \textit{in situ} \cite{2017SSRv..213..219B} and, in particular, to search for the sources of auroral radio emission.
This is made possible by a suite of instruments capable of acquiring high--quality plasma and wave measurements, such as Waves \cite{2017SSRv..213..347K}, JADE--E \cite<Jovian Auroral Distributions Experiment--Electrons,>[]{2017SSRv..213..547M} and MAG \cite{2017SSRv..213...39C}. Imagers on--board Juno are also really useful to compare with auroral emissions in other wavelengths, such as in ultraviolet with the UVS instrument \cite<Ultraviolet Spectrograph>[]{2017SSRv..213..447G}.

These instruments provide measurements to study the radio wave amplification process, and have already been able to locate the position of the sources \cite{2017GeoRL..44.6508I, 2019GeoRL..46..571I, 2019GeoRL..4611606L} and to confirm the Cyclotron Maser Instability (CMI) as their underlying generation mechanism \cite<>[and see below for more details]{2017GeoRL..44.4439L, 2018GeoRL..45.9408L, 2017GeoRL..44.9225L, 2020GeoRL..4790021L, 2023_Collet_PRE9}.

The Galilean moons Io, Europa and Ganymede are known to induce auroral emissions, at radio \cite{1964Natur.203.1008B, 2017JGRA..122.9228L, 2017pre8.conf...45Z, 2018A&A...618A..84Z, 2022A&A...665A..67J}, ultraviolet \cite<UV,>[]{1996Natur.379..323P, 1998...FGH767C, 2002Natur.415..997C}  and infrared \cite{1993Sci...262.1035C, 2017GeoRL..44.5308M, 2018Sci...361..774M} wavelenghts . The motion of the moons across Jupiter's magnetosphere in the plasma torus surrounding them \cite{2022GeoRL..4998111S} generates an electric field, inducing electric currents and/or Alfv\'en waves \cite{1969ApJ...156...59G, 1980JGR....85.1171N, 2004JGRA..109.1210S} which both accelerate electrons along the magnetic field lines in the moons' flux tubes to kilo--electron--volts (keV) energy. Note that the case of Callisto is not studied here, even if a tentative detection of the Callisto UV footprint has been reported \cite{2018JGRA..123..364B}, and hints of radio emissions have been observed to this day using Galileo and Voyager data \cite{2001GeoRL..28.3047M, 2007JGRA..112.5213H}, and not seen by Juno.

The Io, Europa, and Ganymede induced UV emissions are known to be produced by downgoing electrons interacting with the Jovian neutral atmosphere. These signatures are observed at the moons' magnetic footprint and along their tails, i.e the longitudinal extension of these spots in the downstream direction relative to the plasma flow encountering the moon \cite{2017Icar..292..208B, 2017JGRA..122.7985B}. Recent \textit{in situ} studies probed the magnetic field lines connected to these UV footprints, and found that they are consistent with production by Alfv\'enic interaction \cite{2018JGRE..123.3061S, 2020GeoRL..4786527S, 2020GeoRL..4789267S, 2020GeoRL..4789732A}. On the radio side, the moons' induced emissions are believed to be produced by the CMI and have already been simulated and well match the observations \cite{2008GeoRL..3513107H, 2017GeoRL..44.9225L, 2019A&A...627A..30L}. This mechanism is also responsible for the auroral radio emission (independent of the moons) and has been verified \textit{in situ} by \citeA<>[involving loss--cone electron distribution functions, or EDF]{2017GeoRL..44.4439L}, \citeA<>[conics--type EDF]{2018GeoRL..45.9408L} and \citeA<>[shell--type EDF]{2023_Collet_PRE9}.
Recently, \citeA{2020GeoRL..4790021L} showed with \textit{in situ} Juno measurements that the radio emission induced by the Jupiter--Ganymede interaction is indeed produced by the CMI, from a loss cone--type EDF, i.e., a lack in the up--going electron population, with a characteristic energy of $4$--$15$~keV.

Since Jupiter--satellite radio and UV emissions are expected/assumed to be colocated \cite{2010JGRA..115.6205H}, the question of the link between these emissions at two different wavelengths naturally arises. In the Io case, we know that UV and radio auroral emissions are produced by Alfv\'enic interactions, and that the main and secondary radio emissions are respectively produced on the magnetic field lines connected to the main Alfv\'en wing (MAW) and reflected Alfv\'en wing (RAW) spots, and highly suspected for the Transhemispheric Electron Beam (TEB) spots \cite{2010JGRA..115.6205H, 2022JGRA..12730160L}. But no simultaneous \textit{in situ} measurements have yet been analyzed. 
In the Ganymede case, \citeA{2020GeoRL..4790021L} showed, extending the work of \citeA{2020GeoRL..4786527S}, that radio emission is produced above a magnetic flux tube mapping to a UV RAW spot. \citeA{2022GeoRL..4996994H} showed that radio emission seems to be produced above the TEB spot.
Finally in the Europa case, UV emissions have been observed at the moon's footprint and along the moon's footprint tail \cite{2017Icar..292..208B, 2017JGRA..122.7985B, 2020GeoRL..4789732A, 2023JGRA..12831363H, 2023RabiaGRL}, but no simultaneous observation of UV and radio emissions has yet been analyzed.

This study is a follow--up of the \citeA{2020GeoRL..4790021L} analysis, focusing on the three known types of Jupiter--satellite radio emissions. In Section \ref{sec:CMI}, we briefly recall the theory of the Cyclotron Maser Instability. In Section \ref{sec:results}, we present the observations of Jupiter--Io (J--I), --Europa (J--E) and --Ganymede (J--G) radio emission source crossings and calculate the CMI growth rate (whenever possible) and determine the emission parameters. Finally, in Section \ref{sec:sum_disc}, we summarize and discuss the results.

\section{The Cyclotron Maser Instability}
\label{sec:CMI}
The CMI is known to be responsible for the production of auroral radio emission of Earth, Saturn and Jupiter \cite{1979ApJ...230..621W, 1984PhFl...27..247L, 1984JGR....89.2831L, 1985SSRv...41..215W, 1986JGR....9113569P, 2006A&ARv..13..229T, 2010GeoRL..3719105M, 2010GeoRL..3712104L, 2011pre7.conf...75K, 2017GeoRL..44.4439L, 2018GeoRL..45.9408L}.

In a tenuous and magnetized enough plasma, i.e., wherever the electron plasma frequency $f_\text{pe}$ is much lower than the electron cyclotron frequency $f_\text{ce}$, and with weakly out--of--equilibrium/non--maxwellian relativistic electrons, the CMI can directly amplify X--mode waves at a frequency close to the electron cyclotron frequency $f_\text{ce}$, along the surface of a hollow cone. The CMI is a wave--electron instability for which the resonance condition is reached when the Doppler--shifted angular frequency of the wave in the frame of the electrons ($\omega - k_{||} v_{r_{||}}$) is equal to the relativistic gyration frequency of resonant electrons ($\omega_{ce} \Gamma_r^{-1}$): 

\begin{equation}
    \omega = \omega_{ce} \sqrt{1-\frac{v_r^2}{c^2}} + k_{||} v_{r_{||}}~~~~,
\label{eq:resonance_condition}
\end{equation}

with \textbf{k} the wave vector and $v_{r}$ the velocity of the resonant particle, and $\Gamma_r^{-1} = \sqrt{1 - v_r^2/c^2}$ the relativistic Lorentz factor. The $_\perp$ and $_{||}$ indices represent the perpendicular and parallel components of the wave vector\textbf{k} or the velocity $v_{r}$ with respect to the magnetic field $B$. 

In the weakly relativistic case ($v_r \ll c$), the above resonance condition can be rewritten as the equation for a resonant circle in the [$v_\perp$, $v_{||}$] velocity space:
\begin{equation}
	v_{\perp}^2 + (v_{||}  - v_0)^2 = v_{r}^2~~~~,
\label{eq:resonant_circle}
\end{equation}
defined by its center:
\begin{equation}
v_0 = \frac{k_{||}c^2}{\omega_{ce}}~~~~,
\label{eq:resonant_circle_center}
\end{equation} 
and its radius 
\begin{equation}
v_{r} = \sqrt{v_0^2 -2c^2 \Delta \omega} ~~~~,
\label{eq:resonant_circle_radius}
\end{equation}
with 
\begin{equation}
\Delta\omega = (\omega - \omega_{ce}) / \omega_{ce}
\label{eq:delta_omega}
\end{equation}
the frequency shift between the emission frequency and the cyclotron electron frequency.

For the CMI to amplify radio emission, the electron distribution function needs to present a positive $\partial f/\partial v_\perp$ gradient along the resonant circle in the velocity space, and radio waves will be amplified if positive growth rates are presents.

The simplified version of the growth rate expression used by \citeA{2017GeoRL..44.4439L, 2018GeoRL..45.9408L, 2020GeoRL..4790021L} is well adapted to the amplification of X--mode waves propagating at frequencies close to $f_{ce}$, for a refraction index $N=1$ and a moderately energetic ($E \ll 511$~keV) and low--density ($f_{pe} \ll f_{ce}$) plasma. But this expression contains an approximation at low pitch angle, and therefore applies only to growth rate calculation in the loss cone. Therefore to generalize the calculation of growth rate in the whole electron distribution function, we use the expression of \citeA{2023_Collet_PRE9}, derived from the dispersion relation in X mode from \citeA{1984JGR....89.2831L} \cite<see Annexe A of>[for the full demonstration of the growth rate expression]{2023_Collet_PRE9}. They assumed that the plasma is composed of one cold population at thermodynamic equilibrium and one non--thermal energetic (or hot) population. In this study, the hot electron density is the one measured by JADE--E for electrons above $1$~keV energy.

\begin{equation}
	\gamma = \frac{(\frac{\pi}{2} \epsilon_h)^2}{1+(\frac{\epsilon_c}{2\Delta \omega})^2} c^2  \int_0^\pi d \theta v_r^2 \sin^2 (\theta) \frac{\partial f_h}{\partial v_\perp} (v_0 + v_r \cos (\theta), v_r \sin(\theta) )
    \label{eq:growthrate}
\end{equation}

In this equation, $\epsilon_\alpha = \omega_{p \alpha} / \omega_\mathrm{ce}$, where $\omega_{p \alpha}$ is the plasma frequency of the hot ($\alpha =  \mathrm{h}$) or cold ($\alpha = \mathrm{c}$) electrons. $f_h$ is the normalized electron distribution function of hot electrons  ($\int f dv^3 = 1$). In practice, the factor to normalize the distribution function is $c^3 10^{-18}/n_e$, where $n_e$ is the electron density (in cm$^{-3}$).

Equation \ref{eq:growthrate} means that the growth rate is obtained by integrating $\partial f_h / \partial v_\perp$ along a resonant circle in the normalized velocity space [$v_{||}$,$v_\perp$], defined by its center $v_0$ (Equation \ref{eq:resonant_circle_center}), its radius $v_r$ (Equation \ref{eq:resonant_circle_radius}) and the angle $\phi \subseteq$[$0$--$\pi$] along this circle. 

By calculating and maximizing the growth rate, we are able to assess the most CMI--unstable electron population and characterize the resulting amplified waves, and then obtain the characteristics of the emission (e.g., the energy of the resonant electrons and the aperture of the beaming angle).



One of the 3 anodes of JADE--E is unfortunately not functional. As a result, due to Juno's spin and its orientation with respect to the magnetic field lines, it sometimes happens that certain pitch angles are not sampled.
In order to calculate the growth rate of the wave along the different resonance circles in velocity space, JADE--E needs to sample sufficient pitch angles. If too much of the EDF measurement is missing ($60$\% along a resonant circle), we cannot calculate the growth rate and determine the characteristics.

However, assuming that Juno is located within the radio source region (if the radio emission is observed very close to, or even below, $f_{ce}$), we are still able to obtain some information about the source size and the characteristics of the emission, by measuring the emission frequency observed by Juno/Waves data. If the EDF is of shell type, i.e. if $f \le f_\mathrm{ce}$, then the resonant circle is centered on $v_0 = 0$, therefore $k_{||} = 0$ (see Equation \ref{eq:resonant_circle_center}), and Equation \ref{eq:resonance_condition} can then be rewritten as:

\begin{equation}
    \omega_{shell} = \omega_{ce} \sqrt{1-\frac{v_r^2}{c^2}} ~~~~.
\label{eq:resonance_eq_shell}
\end{equation}

Thus, from the measurements of the local electron cyclotron frequency ($\omega_{ce} = 2 \pi f_{ce}$) and the emission frequency $f_\mathrm{shell}$, and using $E= 0.5\times m_e v^2$ (with $m_e = 511$~keV$/\text{c}^2$ the electron mass),  the electron energy in keV in the shell--driven CMI case can be written as:

\begin{equation}
    E \simeq 255.5 \times \left (1- \left( \frac{f_\mathrm{shell}}{f_\mathrm{ce}}\right) ^2\right) ~~~~.
    \label{eq:electron_energy_shell_case}
\end{equation}

In the case of a loss--cone (lc) type EDF, i.e. if $f > f_\mathrm{ce}$, in the weakly relativistic case ($v_r \ll c$) the resonant equation can be rewritten as \cite<for more details see Equations 2--12 of>[]{2019A&A...627A..30L}:

\begin{equation}
    \omega_{lc} \simeq \frac{\omega_{ce}}{\sqrt{1-\frac{v_r^2}{c^2}}} ~~~~,
    \label{eq:resonant_eq_simple}
\end{equation}
and therefore the electron energy in keV in the loss cone--driven CMI case can be written as:

\begin{equation}
    E \simeq 255.5 \times \left (1- \left( \frac{f_{ce}}{f_\mathrm{lc}}\right) ^2\right)
    \label{eq:electron_energy_lc_case}
\end{equation}

\section{Observations and Analysis of Radio Emission Sources Crossings}
\label{sec:results}
Due to the large extension of Io's tail \cite{2020GeoRL..4789267S} and to Juno's polar orbit, the spacecraft crossed Io's magnetic flux tubes at least twice every orbit (North, then South). However, electron fluxes connected to Io's UV aurora are not observed in appreciable or detectable amounts by Juno/JADE--E every transit of these flux tubes. Therefore, during the first 26 Juno perijoves, 18 cases of electron fluxes connected to Io's magnetic flux tube have been reported using the JADE--E measurements \cite{2020GeoRL..4789267S}. By studying the data from perijoves (PJ) \#$27$ to \#$31$, we report five more cases of Io's tail flux tube crossing where electron fluxes were measured. In the Europa case, electron fluxes connected to Europa's UV aurora were measured ten times \cite{2020GeoRL..4789732A, 2023RabiaGRL}. Finally, electron flux connected to Ganymede's UV aurora were measured only two times. The first one during PJ \#$20$ \cite<reported by>[]{2020GeoRL..4786527S, 2020GeoRL..4790021L} and the second one during PJ \#$30$ \cite<studied in details by>{2022GeoRL..4996994H}.

For all moon's flux tube crossings detected by JADE--E, we investigate Waves observations to look for radio emission located below $1$~\% the local electron cyclotron frequency $f_\mathrm{ce}$ (determined from the local magnetic field amplitude measured by the MAG instrument). We therefore considered these cases as a potential crossing of a radio source. We then study the EDF obtained from JADE--E.
\citeA{2020GeoRL..4786527S, 2020GeoRL..4789267S, 2020GeoRL..4789732A, 2023RabiaGRL} studied the downward electrons and the production of UV emissions linked to these downward electron currents, as well as the presence of Alfv\'enic current systems capable of accelerating these electrons. We study here the CMI--unstability of measured EDF, in the continuity of \citeA{2020GeoRL..4790021L}. To go further than \citeA{2020GeoRL..4790021L}, we study instability in the whole EDF, to search not only for loss--cone type instabilities, but also shell type. We also study the upward and downward electrons, as well as the magnetic field perturbation, to determine the presence of field aligned currents (FAC) using \citeA{2021JGRA..12629469W, 2022JGRA..12730586A} method and Alfv\'en perturbations capable of accelerating electrons.

Downtail distances with respect to the main spot were recently revised using Juno/UVS data. Over 1,600 spectral images of the Io, Europa, and Ganymede UV footprint were analyzed to provide statistical positions of the main Alfv\'{e}n wing spots. This allowed \citeA{2023JGRA..12831363H}  to estimate the distance from Juno to the main spot at the time of the source crossings that will be described in this Section, as well as derived observationally an estimation of the Alfv\'{e}n travel time for each three moons. Note that the actual position of the main Alfv\'{e}n wing spots is affected by the background magnetospheric conditions (density of the plasma sheet along the field line connected to the satellite footprint and/or magnetic field strength), and therefore shifts of the main Alfv\'{e}n wing spots mapped to the equatorial region up to $\pm2^{o}$ (Io), $\pm4^{o}$ (Europa) and $\pm5^{o}$ (Ganymede) are not unusual \cite<See>[Figures 4, 5, 7]{2023JGRA..12831363H}. A negative distance to the main spot therefore translate either (i) a source crossing associated with a transhemispheric electron beam located much upstream of the Alfv\'{e}n wing spots, or (ii) a change in the plasma condition (e.g., lower plasmasheet density and/or higher magnetic field magnitude).

\subsection{Jupiter--Io radio emission source crossings}

Out of the 23 cases where electron fluxes connected to Io's tail flux tube were measured, simultaneous radio emissions below $1.01 \times f_\mathrm{ce}$ were observed in only 4 cases. Figure \ref{fig:observation_Io1} displays Juno measurements around an Io flux tube encounter, during PJ\#5 on 2017--03--27 (2017 March 27th) in the Southern hemisphere \cite<already reported by>[]{2019GeoRL..4611606L}. Panel (A) presents the Juno/Waves measurements \cite<in low--resolution mode,>[]{2017SSRv..213..347K} around the perijove (from $\sim -1.5$~h before to $\sim2.5$~h after). Panel (B) is a $5$~min zoom--in of panel (A) using Juno/Waves high--resolution mode. The decreasing solid--black and dashed--black lines in panels (A,B) represent respectively the electron cyclotron frequency $f_{ce}$ and $1.01 \times f_{ce}$. Panels (C)--(E) show the Juno/JADE--E measurements of (C) the electron differential number flux (or intensity), (D) the electron distribution function of upgoing electrons and (E) the partial electron density (where all the energy population $< 0.1$~keV is not accounted for). Figure \ref{fig:observation_Io1}F displays the FAC calculated based on \citeA{2022JGRA..12730586A}'s method (see sections 1.3 and 2 of their SI for more details). This method used the residual magnetic field perturbation $\delta B$, defined as the difference between the Juno/MAG magnetic field measurements and the magnetic field values obtain from the combination of the  \citeA{2018GeoRL..45.2590C} magnetic field and \citeA{1981JGR....86.8370C} current sheet models. The FAC are then calculated from the residual magnetic field perturbation in the azimuthal direction ($\delta B_\phi$).

\begin{figure}
\centering
 \centerline{\includegraphics[width=1.4\linewidth]{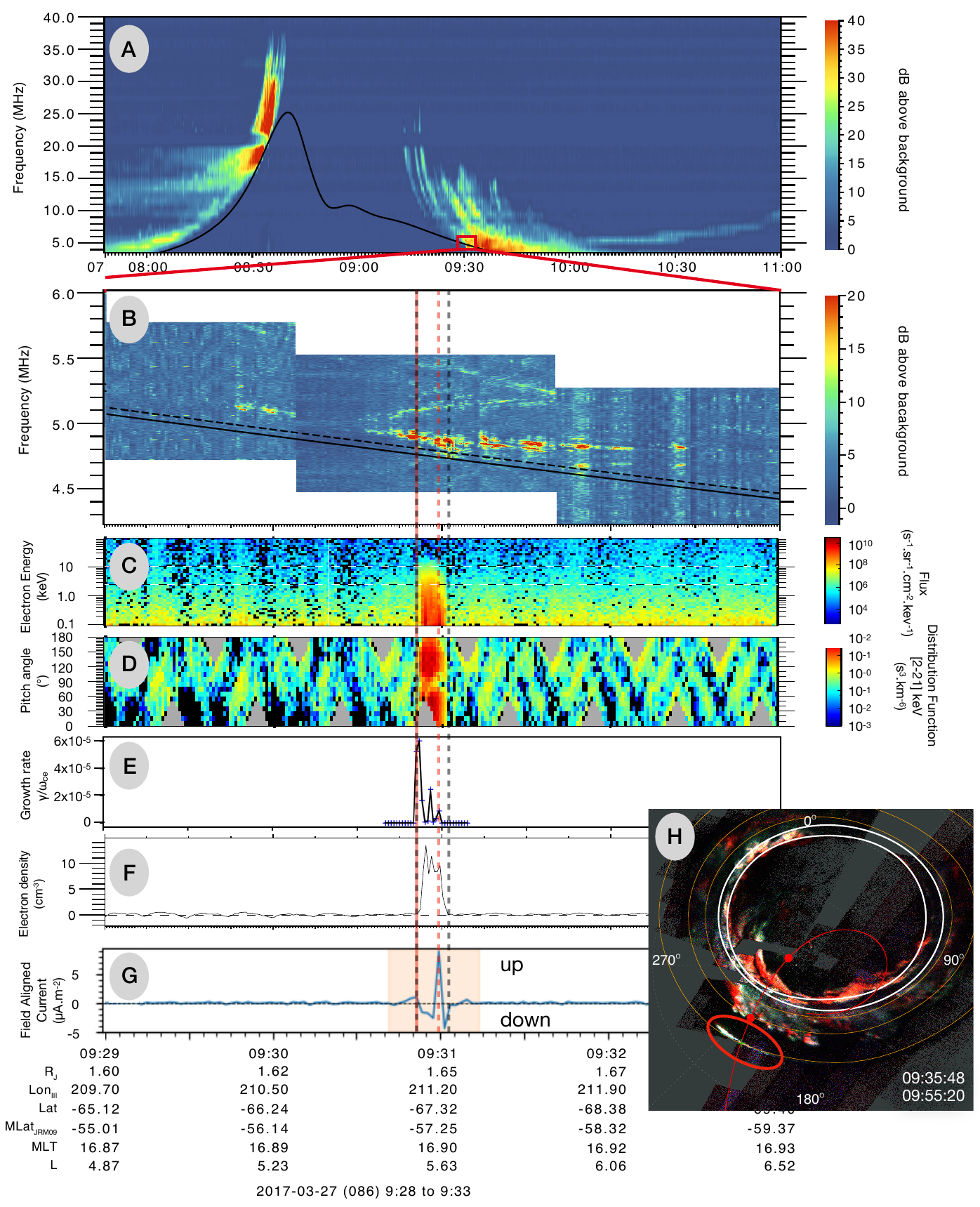}}
\caption{Caption on next page}
\label{fig:observation_Io1}
\end{figure}
\begin{figure}[t]
  \contcaption{Juno data during Perijove 5, on 27 March 2017. Panels (A,B) display Juno/Waves data (A) in low--resolution mode and (B) in high--resolution mode. The solid--black lines represent the electron cyclotron frequency derived from the magnetic field measurements of Juno/MAG, and the dashed--black line is $1.01 \times f_\text{ce}$. Panels (C--D) display Juno/JADE--E measurements: (C) the electron differential number flux (or intensity) of all electrons; (D) the electron distribution function for energy in range [2--21] keV as a function of pitch angles;  Panels (E) displays the normalized growth rate $\gamma / \omega_\mathrm{ce}$ maximal value calculated using Equation \ref{eq:growthrate}. Panel (F) shows the partial electron density calculated from the JADE--E flux. Panel (G) shows the field aligned currents calculated based on \citeA{2022JGRA..12730586A}'s method, using magnetic field fluctuations in the azimuthal direction ($\delta B_\phi$) deduced from the Juno/MAG measurements. The vertical dashed black lines represent the flux tube crossing as inferred from JADE data, while vertical dashed red lines represent the time interval where positive growth rate are calculated from JADE--E measurements. Panel (H) displays a UV map of the southern hemisphere, using Juno/UVS measurements from 09:35:49 to 09:55:20. The red line indicates Juno's trajectory, with the red dots its position at the start and end time of the measurements used for this image. Io UV footprint is highlighted by the red ellipse.}
\end{figure}

Figure \ref{fig:observation_Io1}B displays an emission very close to $1.01 \times f_{ce}$, while we observe an enhancement in the electron energy flux (panel C) at a few keV, a strong intensification in the distribution function (panel D), an increase of the electron density (panel F) and a clear upward current surrounded by downward FAC (panel G). Figure S1 in Supporting Information displays the magnetic field fluctuations for all the magnetic field components in spherical coordinates. The magnetic field perturbation associated to the FAC shows clear fluctuations in the transverse component (perpendicular to B, corresponding to $\delta B_\phi$ and $\delta B_\theta$) while no fluctuations are observed in the radial component ($\delta B_r$). The fluctuations are therefore confined to the transverse components, which is indicative of a lack of horizontal current, and therefore indicative of FAC (displayed Figure \ref{fig:observation_Io1}F). Moreover, no fluctuations are seen in the total magnetic field magnitude $\delta |B|$, indicating that these variations are Alfv\'enic in nature \cite{2019GeoRL..46.7157G, 2019NatAs...3..904K}.

\begin{figure}
    \centering
    \centerline{\includegraphics[width=1.4\linewidth]{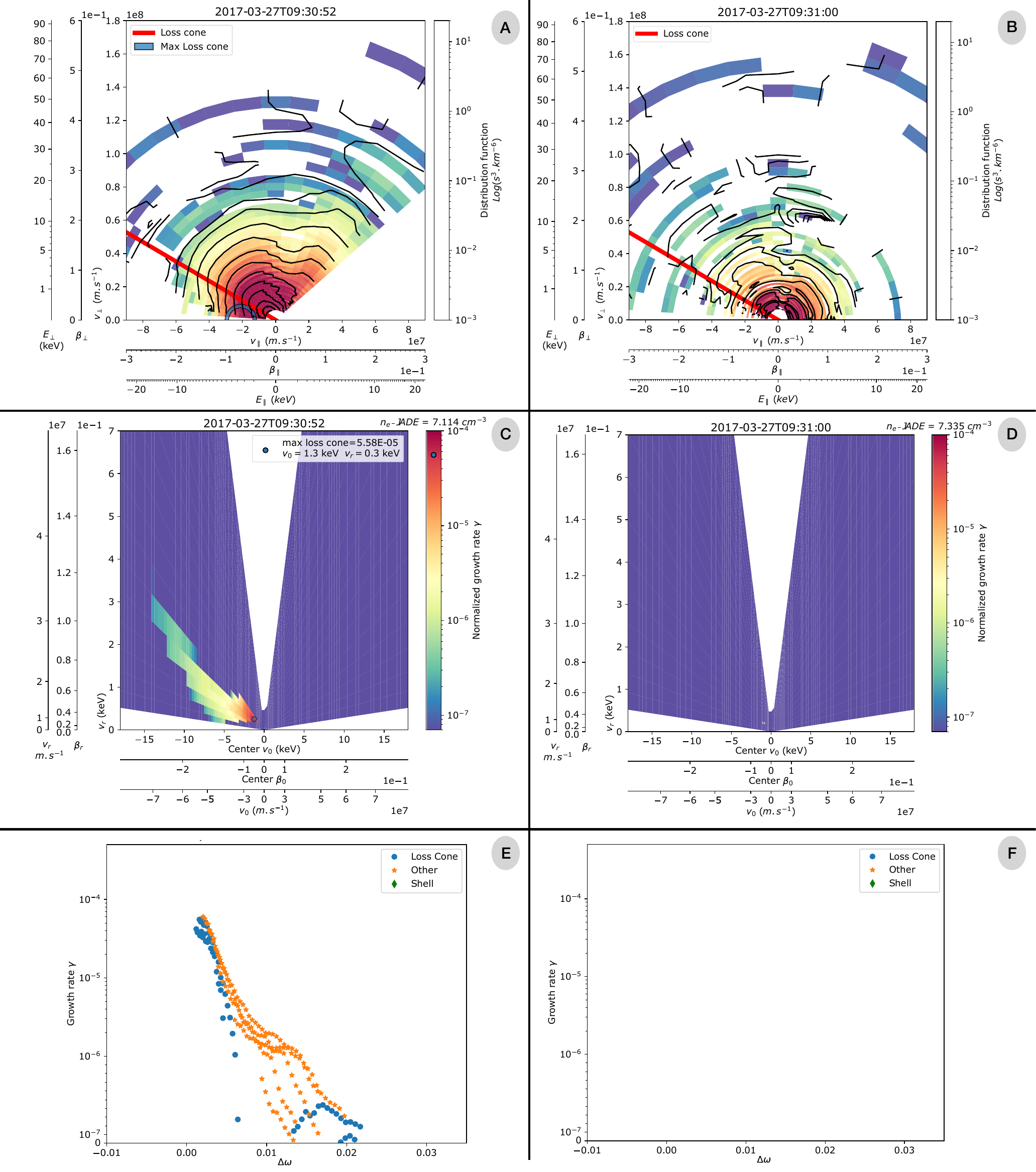}}
    \caption{Caption on next page}
    \label{fig:growthrate_Io1}
    \end{figure}
\begin{figure}[t]
  \contcaption{Panels (A,B): electron distribution function in the velocity space [$v_{||}$, $v_{\perp}$] measured by JADE--E on 2017--03--27 at (A) 09:30:52 (inside the Io tail radio source) and (B) 09:31:00 (outside the Io tail radio source). In that case, the $v_{||} < 0$ part of the EDF represents upgoing electrons, while $v_{||} > 0$ represents downward electrons. The colorbar and the isocontours are shown using a logarithmic scale in units of $s^3.km^{-6}$. The radial red thick line indicates the theoretical loss cone value. The blue circular half--circle in panel (A) display the resonant circle with the highest growth rate.
  Panels (C,D): Normalized growth rate ($\gamma / \omega_{ce}$) estimates for different resonant circles at different centers $v_{0}$ and radii $v_r$. 
  Panels (E,F): Normalized growth rate as a function of the frequency shift $\Delta \omega$ between the emission frequency and the cyclotron electron frequency (see Equation \ref{eq:delta_omega} for all resonant circle with positive growth rate $\gamma$. Blue circles represent growth rate for resonant circles tangential to the theoretical value of the loss cone. Orange stars represent growth rate growth rate for resonant circle inside the theoretical loss cone. Both are considered as loss--cone type instabilities. Green diamonds represent growth rate for resonant circles of shell-type.}
\end{figure}

Figures \ref{fig:growthrate_Io1}A--B display the electron distribution function in the velocity space measured by Juno/JADE--E at (A) 09:30:52 and (B) 09:31:00. From these data and Equation \ref{eq:growthrate} we can calculate the normalized growth rate of the emission along different resonant circles to determine the unstable electron population. Figures \ref{fig:growthrate_Io1}C--D show the estimated normalized growth rate $\gamma/\omega_{ce}$ along resonant circles in the whole EDF, calculated for different centers (x--axis) and different radii (y--axis). Figures \ref{fig:growthrate_Io1}E--F displays the growth rate $\gamma$ as a function $\Delta \omega$ for each resonant circle displaying a positive growth rate.

Positive growth rates are obtained for the EDF of Figure \ref{fig:growthrate_Io1}A, and not for the EDF of Figure \ref{fig:growthrate_Io1}B. Only resonant circles inside the theoretical loss cone show positive growth rate (blue circles and orange stars in Figure \ref{fig:growthrate_Io1}E), while no positive growth rate are found for shell--type resonant circles (green diamonds). By doing this calculation before, during and after the crossing of the Io's tail flux tube (see Figure \ref{fig:observation_Io1}E), we are able to determine when Juno is in the source, and thus determine its size and the characteristics of the emission. In this case, positive growth rate are obtained along loss--cone type resonant circles from 09:30:51 to 09:30:59 ($\pm 1$~sec). This time interval is indicated in Figure \ref{fig:observation_Io1} by the two vertical dashed red lines. Juno's velocity being $\sim 45$~km/s during this time, we determined that the source size is $360 \pm 45$~km. From the growth rate calculation, we can determine that the energy of the resonant electrons responsible for this emission is in the range [$1$--$15$]~keV, with an opening angle $\theta$ of the beaming cone in the range [$74 \degree$--$85\degree$]. To determine this value, we used Equation 7 of \citeA{2020GeoRL..4790021L}:
\begin{equation}
\theta = \text{acos}(\beta_0/(1+\Delta \omega))~~~~, 
\label{eq:delta_omega}
\end{equation}
based on the assumptions of Section \ref{sec:CMI}.

It is interesting to note that enhanced electron fluxes are observed for the period of 09:30:51 to 09:31:02 \cite<determined as the flux tube crossing,>[]{2020GeoRL..4789267S}, while the J--I radio source is only crossed from 09:30:51 to 09:30:59 (determined from the growth rate calculation), which corresponds to the time where Juno is inside a downward current (corresponding to upward electrons). When Juno is located inside an upward current (i.e., downward electrons), no positive growth rate are obtained.

The same method is applied to the data from PJ\#6 (North) on 2017--05--19, during which a J--I radio source is crossed between 05:39:31 and 05:39:39 (see Figure S2), and to PJ\#29 (North) on 2020--09--16 during which a J--I radio source is crossed between 02:00:34 and 02:00:36 (see Figure S3). For the crossing of the J--I radio source that occurs in the northern hemisphere during PJ\#5 on 2017--03--26 around 08:34:40 (see Figure S4), we do not have JADE--E measurements of the upgoing electrons, and we therefore can not calculate the growth rate. Since no radio emission is observed below $f_{ce}$, we therefore assume that loss--cone EDF remain the prominent source of free energy for the CMI, and we therefore use Equation \ref{eq:electron_energy_lc_case} to determine the electron energy. The results for these three crossings, i.e., radio source size, resonant electron energy, $\text{f}_\text{emission}$ and opening angle of the beaming cone, are summarized Table \ref{tab:results}.

As for PJ\#5 (South), FAC and Alfv\'enic perturbations are observed during PJ\#5 (North) and PJ\#29 (North) Io's flux tube crossing (see Figures S4 and S3, respectively), with perturbations in the transverse components of the magnetic field ($\delta B_\phi$ and $\delta B_\theta$) while no perturbation is observed in the radial ($\delta B_r$) and compressive ($\delta |B|$) components. Furthermore, as for PJ\#5 (South), the radio source is not crossed anywhere inside Io's flux tube, but only when Juno is located inside a downward current (i.e., upward electrons). During  PJ\#6 (North), nothing is observed in the magnetic field perturbations, which could be due to the fact that the electron density is very low ($< 1$~cm$^{-1}$), which could induce a perturbation too weak for the MAG instrument to detect.

Based on the recalculation of the downtail distance to the main spot of Io $\Delta \lambda_\mathrm{Alfv\acute{e}n}$ \cite{2023JGRA..12831363H}, the J--I radio emission source crossings of PJ\#5 North, PJ\#5 South, PJ\#6 South are all associated with a Reflected Aflv\'en Wing (RAW) spot downtail of Io. The intensity of the radio emission seems to be quite similar for crossings occurring close to the main spot, with an intensity of $2$--$3 \times 10^{-6}$~$V^2.m^{-2}.Hz^{-1}$ when $ 3.3^{\mathrm{o}} < \Delta \lambda_\mathrm{Alfv\acute{e}n} < 10.8^{\mathrm{o}}$. The intensity seems to be lower when $\Delta \lambda_\mathrm{Alfv\acute{e}n}$ is large, with an intensity of $8 \times 10^{-8}$~$V^2.m^{-2}.Hz^{-1}$ for $\Delta \lambda_\mathrm{Alfv\acute{e}n} = 87.4^{\mathrm{o}}$.

\subsection{Jupiter--Europa radio emission}
\label{sec:Europa}

During the 26 first perijoves, enhanced electron fluxes connected to Europa's UV aurora were measured ten times \cite{2020GeoRL..4789732A, 2023RabiaGRL}.  A radio source was crossed only during PJ\#12 on the Northern hemisphere, on 2018--04--01 around 08:15:44.

\begin{figure}
\centering
 \centerline{\includegraphics[width=1\linewidth]{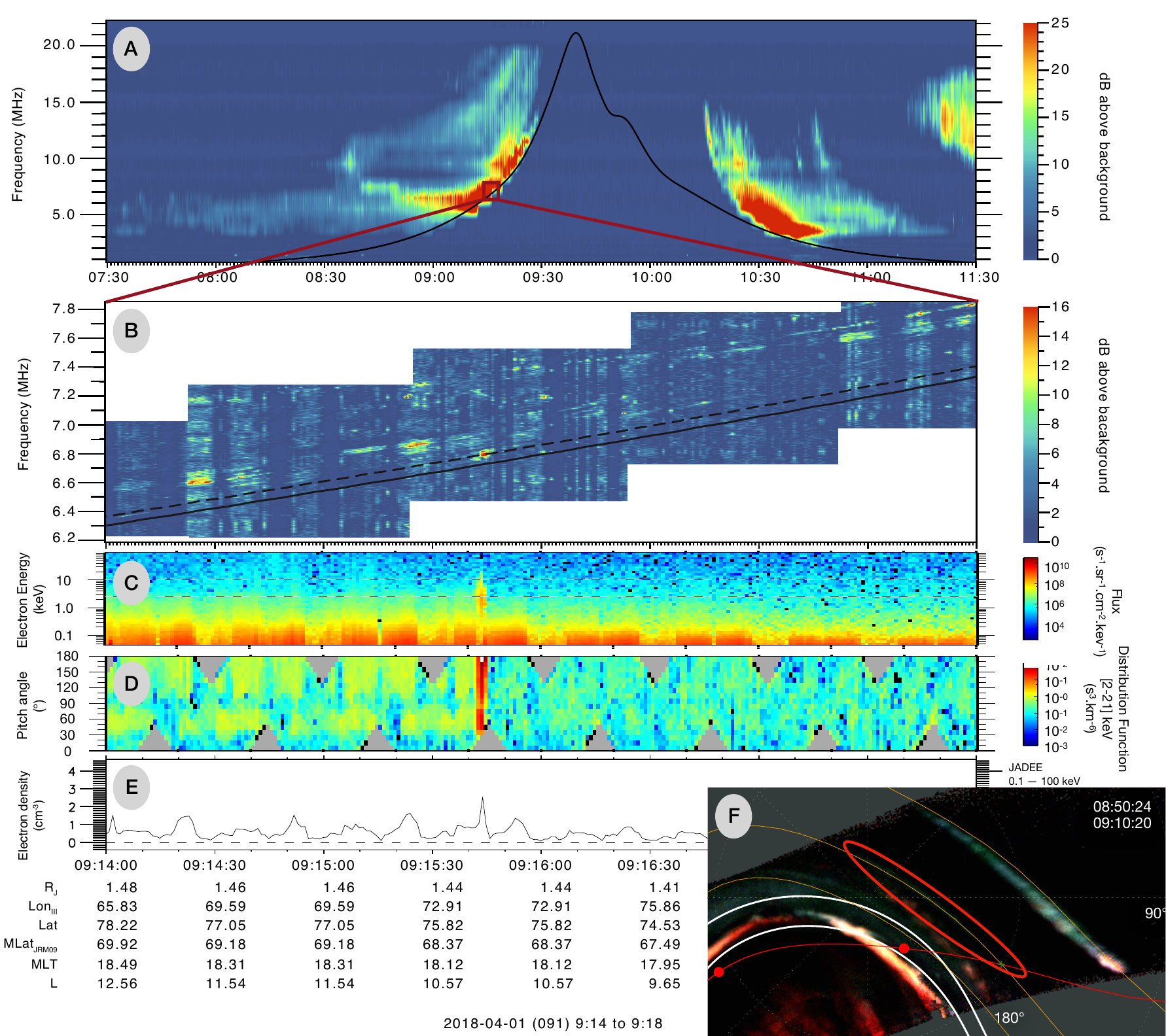}}
\caption{Panels (A,B) display Juno/Waves data (A) in low--resolution mode and (B) in high--resolution mode. The solid--black lines represent the electron cyclotron frequency derived from the magnetic field measurements of Juno/MAG, and the dashed--black line is $1.01 \times f_\text{ce}$. Panels (C--E) display Juno/JADE--E measurements: (C) the electron differential number flux (or intensity) of all electrons; (D) the electron distribution function for energy in range [2--21] keV only for pitch angles [$0\degree$--$60\degree$] corresponding to up--going electrons; (E) partial electron density calculated from the JADE--E flux. Panel (F) displays a UV map of the northern hemisphere, using Juno/UVS measurements from 08:50:24 to 09:10:20. The red line indicates Juno's trajectory, with the red dots its position at the start and end time of the measurements used for this image. Europa UV footprint is highlighted by the red ellipse.}
\label{fig:observation_Europa}
\end{figure}

Figure \ref{fig:observation_Europa} displays Juno/Waves (panels A--B), Juno/JADE--E measurements (panels C--E) and Juno/UVS (panel F) during the crossing of flux tube connected to a Europa's downtail UV footprint \cite{2020GeoRL..4789732A}. During the time of the flux tube crossing, determined by the enhancement of the electron flux in JADE--E measurements (Figures \ref{fig:observation_Europa}C--D), a radio emission is observed below $1.01\times f_{ce}$ (Figure \ref{fig:observation_Europa}B). However, the JADE--E instruments did not record data of upgoing electrons in the loss cone during this time. We therefore cannot calculate the loss--cone or shell CMI growth rate for this EDF. Since no radio emission is observed below $f_\mathrm{ce}$, we therefore assume that loss--cone EDF remain the prominent source of free energy for the CMI. We then apply Equation \ref{eq:electron_energy_lc_case} to estimate the energy of the resonant electron. Looking at the Juno/Waves measurements (Figure \ref{fig:observation_Europa}B), we determine that the radio emission observed during Europa's flux tube crossing is emitted at a frequency between 0.7 and 1.5\% above the electron cyclotron frequency $f_{ce}$ (solid dark line). This frequency measurement leads to an energy of the resonant electrons in the range [$3$--$8$]~keV, and an opening angle of the beaming cone in the range [$79\degree$--$84 \degree$].

The MAG measurements of the magnetic field perturbation show no strong variation of the different components. Once again, as in the case of PJ\#6N for Io, the electron density is very low ($\sim 2$~cm$^{-1}$) which could induce perturbations too weak and/or too short for the MAG instrument to detect.

Based on the latest work of \citeA{2023RabiaGRL} and the recalculation of $\Delta \lambda_\mathrm{Alfv\acute{e}n}$ \cite{2023JGRA..12831363H}, we can conclude that the J--E radio source crossed during PJ\#12 is associated with a Transhemispheric Electron Beam (TEB) spot uptail of the main Europa UV spot.


\subsection{Jupiter--Ganymede radio emission}
\label{sec:Ganymede}

So far, flux tubes connected to Ganymede footprint tail aurora have been crossed twice: the first one during PJ \#$20$ on 2019--05--19 between 07:37:14 and 07:37:32 \cite<reported by>[]{2020GeoRL..4786527S, 2020GeoRL..4790021L} and the second one during PJ \#$30$, on 2020--11--08 around 02:55:02 \cite{2022GeoRL..4996994H}.

We already reported the PJ\#20N crossing in \citeA{2020GeoRL..4790021L}, but at that time, we did not look at the Juno/MAG measurements, which was done by \citeA{2020GeoRL..4786527S} (and plotted in Figure S5). During this Ganymede footprint tail aurora flux tube crossing, fluctuations in the transverse components ($\delta B_\phi$ and $\delta B_\theta)$ were observed, while no fluctuations were measured in the radial ($\delta B_r$) and compressive ($\delta |B|$) components, which indicates the presence of field aligned currents and Alfv\'enic perturbations.
As for the J--I radio emission sources, it should be noted that the radio source is only crossed when Juno is in a downward current (i.e., upward electrons). Based on \citeA{2020GeoRL..4786527S, 2020GeoRL..4790021L}, this radio emissions is associated with a RAW UV spot (with a  $\Delta \lambda_\mathrm{Alfv\acute{e}n} = 8 \degree$). However, based on the recent work of \citeA{2023JGRA..12831363H}, for this J--G radio source, $\Delta \lambda_\mathrm{Alfv\acute{e}n} = -1.8 \degree$. Therefore, due to the error of $±5 \degree$ on the position of the MAW for Ganymede (due to possible change in the \textit{in situ} plasma condition, see penultimate paragraph of Section \ref{sec:intro}), it appears that the J--G radio source crossed during PJ\#20, is connected not to the RAW spot, but to the MAW spot.

Concerning the second case, during PJ\#30 on 2020--11--08 (northern hemisphere), Ganymede footprint tail aurora flux tube was crossed around 02:55:02, with radio emission tangent to $1.01 \times f_{ce}$ at the same time \cite<see>[]{2022GeoRL..4996994H}. Unfortunately, JADE--E did not measure the upward electrons during Ganymede flux tube crossing. Therefore, we can only estimate the electron energy using Equation \ref{eq:electron_energy_lc_case} (as no radio emission is observed below $f_{ce}$). Around 02:55:02, Waves measured a radio emission between $1.804$~MHz and $1.894$~MHz. Based on the Juno/MAG measurements of the magnetic field amplitude, $f_{ce} = 1.7857$~MHz. These values of the emission frequency lead to an estimation of the resonant electron energy in the range [$5.1$--$28.5$]~keV, and an aperture angle of the beaming cone in the range [$70 \degree$--$81\degree$], provided that Juno actually flew through the radio source. During the crossing, MAG measurements show a $\sim 10$~nT perturbation in $\delta B_\phi$, while no perturbation is observed in the radial $\delta B_r$ and compressive $\delta |B|$ components, which indicates again the presence of Alfv\'enic perturbations and field aligned currents, with an upward current equatorward of a downward current. 

Based on the work of \citeA{2022GeoRL..4996994H}, we can also conclude that the J--G radio emission source crossed during PJ\#30 is associated with a Transhemispheric Electron Beam (TEB) spot uptail of the main Europa UV spot.

The results for these two J--G radio source crossings are summarized in Table \ref{tab:results}.

\section{Summary and Discussion }
\label{sec:sum_disc}

\begin{table}[h]
\scriptsize
\centering
\caption{Results for the Jupiter--Io, Jupiter--Europa and Jupiter--Ganymede radio emissions source crossings. Is given for each crossing:
the name of the moon;
the hemisphere of the emission;
the associated perijoves;
the date and time interval of the radio source crossing as inferred from growth rate calculation when JADE data were available;
the JADE data availability; 
the minimal frequency reached by the radio emission (in MHz);
the frequency bandwidth of the emission (in percentage above $f_\mathrm{ce}$); 
the maximum intensity (in V$^2$.m$^{-2}$.Hz$^{-1}$) of the emission;
the electron energy (in keV);
 the opening half--angle of the beaming cone (in $\degree$);  
the radio source size (in km);
the downtail distance to the Main Alfv\'en Wing spot $\Delta \lambda_\mathrm{Alfv\acute{e}n}$ \cite{2023JGRA..12831363H};
the associated UV emission at the footprint of the magnetic field line associated to the source (MAW: Main Alfv\'{e}n Wing: RAW: Reflected Alfv\'en Wing; TEB: Transhemispheric Electron Beam).}
\resizebox{1.1\textwidth}{!}{\colortable{%
\begin{tabular}{c|c|c|c|c|c|c|c}
\toprule
Moon                            & Io & Io & Io  & Io & Europa & Ganymede & Ganymede \\ 
Hemisphere & South  & North & North & North & North & North & South   \\ 
Perijove                            & PJ5                         & PJ5         & PJ6         & PJ29           & PJ12  & PJ20         & PJ30     \\ 
Date (Year--Month--Day)                            & 2017--03--27 & 2017--03--27 & 2017--05--19 & 2020--09--16 & 2018--04--01 & 2019--05--29  & 2020--11--08 \\
Time interval (HH:MM:SS)                            & 09:30:51--59 & around 08:34:40 & 05:39:31--39 & 02:00:34--36 & around 09:15:44 & 07:37:25--30  & around 02:55:02 \\
                            
\midrule                
JADE data                   & Yes                           & No            &  Yes      & Yes   & No & Yes & No\\
$f_\mathrm{min}$ (MHz) & $4.7$ & $20.8$ & $12.8$ & $27.7$  & $6.7$ & $6.5$ & $1.8$  \\
$f_\text{emission}$ (\% $>f_{ce}$) & $3$--$18\times10^{-3}$    & $3$--$29\times10^{-3}$ & $2$--$14\times10^{-3}$ & $5$--$40\times10^{-3}$  & $7$--$15\times10^{-3}$ & $5$--$21\times10^{-3}$ & $5$--$40\times10^{-3}$\\
Intensity max. (V$^2$.m$^{-2}$.Hz$^{-1}$) & $3\times10^{-6}$ &  $3\times10^{-6}$ & $8\times10^{-8}$ & $2\times10^{-6}$ & $1\times10^{-7}$ & $1\times10^{-6}$ & $3.5\times10^{-9}$ \\
Electron energy (keV)       & $1$--$15$                      & $2$--$20$      &  $1$--$5$ & $3$--$10$  & $3$--$8$ & $4$--$15$ & $2$--$20$\\
Opening angle               & $74$--$85\degree$ & $74$--$85\degree$ & $77$--$86\degree$ & $73$--$84\degree$ & $79$--$84\degree$ & $76$--$83\degree$ & $74$--$85\degree$  \\
Radio source size (km)      & $360 \pm 45$                  & $500\pm 100$  & $415 \pm 50$ & $250 \pm 50$  & $200 \pm 49$  & $250 \pm 50$ & $75 \pm 50$\\
$\Delta \lambda_\mathrm{Alfv\acute{e}n}$ ($^{o}$) & $3.3^{\mathrm{o}}$ & $10.8^{\mathrm{o}}$ & $87.4^{\mathrm{o}}$ & $7.8^{\mathrm{o}}$ & $-10.5^{\mathrm{o}}$  & $-1.8^{\mathrm{o}}$ & $-7^{\mathrm{o}}$ \\
Associated UV emission & RAW & RAW & RAW & RAW  & TEB & MAW & TEB \\
\bottomrule
\end{tabular} 
}}
\label{tab:results}
\end{table}

Concerning the characteristics of the radio emission, the results are similar for the three Galilean moons Io, Europa and Ganymede, in terms of driving mechanism (CMI), electron energy, and beaming. The \textit{in situ} measurements by JADE--E show that the radio emission is triggered by the Cyclotron Maser Instability, driven by a loss--cone electron distribution function. No unstable shell--type electron distribution function are detected in JADE--E measurements. The energy of the resonant electrons is in the range [$1$--$20$]~keV, and the half--opening cone angle is in the range [$74\degree$--$86\degree$]. 

These values are in agreement with those recently obtained using ground--based radio observation, such as the Nan\c{c}ay Decameter Array or NenuFAR, such as the recent work of \citeA{2022JGRA..12730160L} who determined for Io an opening angle $\theta(f)$ in the range [$70\degree$--$80\degree$] and electron energies in the range [$3$--$16$]~keV. For Europa, \citeA{2023PREIX_Lamy} measured on an unique detection an opening angle in the range $\theta$ = [$80\degree$--$86\degree$] and an electron energy in the range [$0.5$--$3$]~keV. For Ganymede, the observations of three emissions lead them to a determination of a beaming angle in the range $\theta$ = [$71\degree$--$87\degree$] and an electron energy in the range [$0.5$--$15$]~keV

The radio sources have a latitudinal extent of a few hundreds of kilometers. In the cases where we are able to constrain the radio source location (provided that we have JADE--E measurement of the up--going electrons), the sources were not crossed anywhere in the flux tube, but only in the downward Field Aligned Currents.

Based on the previous works of \citeA{2020GeoRL..4786527S, 2020GeoRL..4789267S, 2020GeoRL..4790021L, 2022GeoRL..4996994H, 2023RabiaGRL} and the recalculated downtail distances from the UV moon main spot using \citeA{2023JGRA..12831363H}, we also concluded that in the case of Io, all radio source crossed were associated with a RAW UV spot. These crossed radio sources are therefore associated with the secondary radio emissions observed in the usual dynamic spectrum.

In the case of Europa, the only case of radio emission source so far is associated with a TEB spot. Finally, for Ganymede, one radio source is associated with a MAW spot, while the second one is associated with a TEB spot. Even if we didn't detect any radio emission above TEB for Io, these results are is in agreement with the interpretation of the first identification of some Io--DAM linked to the TEB spot analyzed in \citeA{2022JGRA..12730160L}.

The maximal intensity is quite similar between all cases close to the main UV spot, with a value of $2$--$3 \times 10^{-6}$~$V^2.m^{-2}.Hz^{-1}$ in a interval $3.3^{\mathrm{o}} < \Delta \lambda_\mathrm{Alfv\acute{e}n} < 10.8 ^{\mathrm{o}}$, with a decrease of the intensity with long distance downtail ($8 \times 10^{-8}$~$V^2.m^{-2}.Hz^{-1}$ for the case at $\Delta \lambda_\mathrm{Alfv\acute{e}n} = 87.4^{\mathrm{o}}$). But with only one case very far downtail, we can't produce a fit of this decrease as a function of $\Delta \lambda_\mathrm{Alfv\acute{e}n}$. However, the maximal intensity of the emission is quite smaller for the radio source crossed above a TEB spot ($10^{-9}$--$10^{-7}$~$V^2.m^{-2}.Hz^{-1}$). This could be related to the type of electron distribution, which seems different near tail (non--monotonic) than far tail (broadband), at least for Europa with a separation at $ \Delta \lambda_\mathrm{Alfv\acute{e}n} \simeq 4^{\mathrm{o}}$ downtail to the Main spot \cite{2023RabiaGRL}.

From these latest observations, it therefore appears that the cyclotron maser instability driven by a loss--cone electron distribution function is a common way of amplifying radio emission at Jupiter. This is the case both for auroral radio emission \cite{2017GeoRL..44.4439L, 2018GeoRL..45.9408L} and for moon--induced radio emission \cite<>[and this present study]{2020GeoRL..4790021L, 2022GeoRL..4996994H}. We have also shown here that the presence of Alfv\'enic perturbation as well as field aligned current are necessary for the radio emissions to be amplified. The radio sources are located only in the downward part of the FAC, i.e. when the current is carried by upgoing electrons. This supports the results obtained from very high resolution observations \cite{1996GeoRL..23..125Z, 2004P&SS...52.1455Z, 2007P&SS...55...89H, 2007JGRA..11211212H, 2022RASTI...1...48L}, which show that the millisecond bursts observed in the J--I emissions present only negative drifts, i.e. upward--moving electrons. Finally, radio emission are found to be associated with TEB, MAW and RAW spot at the footprint of the flux tube connected to the moons.

However, the Cyclotron Maser Instability does not trigger radio emission at a detectable level for Waves every time Juno is in the flux tube of the moons, even if UV emission is observed at the footprint of the flux tube in each case. Radio sources are crossed in the two cases of Ganymede flux tube crossings. In contrast, a radio source is crossed only once over ten Europa flux tube crossings, while for Io, only four radio sources are crossed over 23 Io flux tube crossings. Therefore, it is clear that several criteria are necessary to amplify a radio wave through the CMI.  

First, we knew that for the CMI to occur, a low energetic plasma is needed ($f_{pe}/f_{ce} \ll 0.1$). But in this study, we also found that the CMI needs to have a sufficiently dense, hot and energetic plasma to occur. If the ratio between $f_{pe}$ and $f_{ce}$ is too low, the integration of the $\delta f / \delta v_{\perp}$ gradient along the resonant circles gives an insufficiently high growth rate to amplify the wave to an observable intensity. 

A second necessary condition seems to be the presence of an Alfv\'enic acceleration process and Field Aligned Current. Could the loss--cone--driven CMI be triggered by upward electrons accelerated by a Fermi acceleration process in the FAC generated by the Alfv\'en Waves \cite<as suggested by>[]{1997JGR...102...37C}. To answer this question, more crossings of Jupiter-Moon radio emissions will be necessary, and future Juno observations could further illuminate these important processes.

\section*{Data Availability Statement}
The Juno data used in this manuscript are found at the Planetary Data System at \url{https://doi.org/10.17189/1522461} for Waves data \cite{2022_waves_survey_Kurth}, at \url{https://doi.org/10.17189/1519715} for JADE--E data \cite{2022_JADEE_pds_Allegrini} and at \url{https://doi.org/10.17189/1519711} for MAG data \cite{2017_FGM_pds_Connerney}.

\acknowledgments
The authors thank the Juno mission team, especially the staff of the Waves, JADE and MAG instruments.
C. K. L.'s work at IRAP was supported by CNES.
CKL's and CMJ's work at DIAS was supported by Science Foundation Ireland Grant 18/FRL/6199.
C. K. L. thanks E. Penou, for the help with the particle analysis through the CLWeb software. 
V. H. acknowledges support from the French government under the France 2030 investment plan, as part of the Initiative d'Excellence d'Aix--Marseille Universit\'{e} -- A*MIDEX AMX--22--CPJ--04.
The research at the University of Iowa is supported by NASA through Contract 699041X with the Southwest Research Institute. 
The research at the Southwest Research Institute is supported by NASA New Frontiers Program for Juno through contract NNM06AA75C.
The french authors acknowledge support from CNES and CNRS/INSU national programs of planetology (PNP) and heliophysics (PNST).

\bibliography{bibliography.bib}

\end{document}